\documentclass{llncs}  
\newtheorem{defn}{D\'efinition} [section]
\newtheorem{thm}{Theorem}
\newtheorem{lem}[defn]{Lemma}

\title{Some new Features and Algorithms for the Study of DFA}
\author{A.N. Trahtman}
\date{}
\institute{Bar-Ilan University, Dep. of Math. and CS, 52900,Ramat
Gan, Israel  email:trakht@macs.biu.ac.il}

\begin{document}

\maketitle

\centerline{Open Journal of Discrete Mathematics, 2012, 2, 45-50}

\begin{abstract}
 The work presents some new algorithms realized recently in the package TESTAS.
They decide whether or not deterministic finite automaton ($DFA$) is synchronizing,
several procedures
find relatively short synchronizing words and a synchronizing word of the minimal
length. We check the existence of a coloring of directed graph that turns
the graph into a synchronizing $DFA$.
The algorithm finds the coloring (better known as the road coloring) if it exists.
Otherwise, the $k$-synchronizing road coloring can be found.
We use a linear visualization of the graph of an automaton based on its
structural properties.
\end{abstract}


\section*{Introduction}
The problem of synchronization of a DFA is natural and various
aspects of this problem were touched upon the literature.
 Synchronization makes the behavior of an automaton resistant
against input errors since, after detection of an error,
synchronizing word resets the automaton back to its original
state, as if no error had occurred  \cite{BCKP}.
Synchronizing word stops propagation of errors in prefix code.

The early version of the package TESTAS was described in \cite{Tp} in 2003.
There exists some interest in the original algorithms of the package,
sometimes even quite exotic \cite{BOBB}. The features of the package are
considered also  favorably for educational purposes: "The Road Coloring
Conjecture makes a nice supplement to any discrete mathematics course" \cite{Ra}.

A problem with a long story is the estimation of the minimal length of
a synchronizing word, (\v{C}erny's conjecture).
  Jan \v{C}erny found in 1964 \cite {Ce} $n$-state complete DFA with
shortest synchronizing word of length $(n-1)^2$ for alphabet size $q=2$.
 The problem can be reduced to automata with strongly connected graph.
  \v{C}erny's conjecture together with the road coloring problem belong
to the most fascinating problems in the theory of finite automata
\cite{CKK}, \cite{MS}, \cite{Pi}.

The package decides whether or not $DFA$ is synchronizing, several procedures
 find relatively short synchronizing words ($O(n^3d)$ time complexity in the
worst case) and a synchronizing word of the minimal length
(non-polynomial algorithm) \cite{Tk}. The space
complexity is quadratic. These procedures were successfully
checked, in particular, in the program that has studied all
transition graphs of automata with 10 states or less in a search
of long synchronizing words. The size of the set of
studied objects was about $10^{20}$.

Imagine a map with roads which are colored in such a way that
fixed sequence of colors, called a synchronizing sequence, leads
to fixed place whatever is the starting point. Finding such a
coloring is called {\em road coloring problem}. The roads of the map are
considered as edges of a directed graph.

The road coloring conjecture  \cite{AGW}, \cite{AW},  \cite{Ru}
 was stated over forty yeas ago for a complete strongly connected directed
finite graph with constant outdegree of all its vertices where the greatest
 common divisor (gcd) of lengths of all its cycles is one.
The edges of the graph being unlabelled, the task is to find a
labelling that turns the graph into a deterministic finite
automaton possessing a synchronizing word.

The problem was mentioned in "Wikipedia" on the list of the
 interesting unsolved problems in mathematics many years ago.
The positive solution of the road coloring problem  \cite{Tc}, \cite{Ti}
 is a basis of a polynomial-time implemented algorithm of $O(n^3)$ complexity
in the worst case. The space complexity is quadratic.

For arbitrary complete graph the program finds $k$-synchronizing
(or generalized \cite{BF}) road coloring \cite{Tk}.

The visualization of the transition graph of an automaton is an important tool
for the study of automata. A tool for the visualization of the inner structure
of a digraph is without any doubt an interesting matter, not only for the road
coloring problem but also for a wide range of applications on directed graphs
with labels on edges. For these reasons, the visual perception of the structural
properties of an automata is important.

The visualization algorithm is linear in the size of the
automaton \cite{TBC}. This feature of the package is handy.

\section{Preliminary}

As usual, we regard a directed graph (digraph) with letters assigned to its edges as
a finite automaton, whose input alphabet $\Sigma$ consists of these letters.
The graph is called {\it a transition graph} of the automaton. The letters from
$\Sigma$ can be considered as colors and the assigning of colors to edges will
be called {\it coloring}.

A {\it path} in a digraph $G$ is a sequence of edges $e_1,...,e_k$ such that
the end vertex of $e_i$ is the start vertex of $e_{i+1}$ for $ i=1,2,...,k-1$.
The path is called a {\it cycle} if $e_1=e_k$.

A digraph is strongly connected if for every pair of vertices
${\bf q}$, ${\bf p}$ there exists a path from ${\bf q}$ to ${\bf p}$.
An arbitrary digraph consists of some strongly connected components ($SCC$).
An $SCC$ is sink if from every vertex of digraph there exists a path to
vertex of the $SCC$.

A finite directed strongly connected graph with constant
outdegree of all its vertices where the gcd of lengths of all
 its cycles is one will be called an {\it $AGW$ graph} (as introduced by Adler,
Goodwyn and Weiss).

An automaton is {\it deterministic} if no state has two outgoing edges
of the same color. In {\it complete} automaton each state has outgoing
edges of any color.

If there exists a path in an automaton from the state $\bf p$ to
the state $\bf q$ and the edges of the path are consecutively
labelled by $\sigma_1, ..., \sigma_k$, then for
$s=\sigma_1...\sigma_k \in \Sigma^+$ we shall write  ${\bf q}={\bf p}s$.

Let $Ps$ be the set of states ${\bf p}s$ for ${\bf p} \in P$
 $s \in \Sigma^+$. For the transition graph
$\Gamma$ of an automaton let $\Gamma s$
denote the map of the set of states of the automaton.

 A word $s \in \Sigma^+ $ is called a {\it synchronizing}
word of the automaton with transition graph $\Gamma$
if $|\Gamma s|=1$.

 A coloring of a directed finite graph is {\it synchronizing} if the
coloring turns the graph into a deterministic finite automaton
possessing a synchronizing word.

Let the integer $q$ denote the size of alphabet and let $n$
be the number of nodes.

\section{Algorithms for finding synchronizing word}
The package TESTAS presents three distinct versions of polynomial time
algorithm for synchronizing word based on different approaches
\cite{Ts}, \cite{Tk}. The algorithms have $O(n^3d)$ time
complexity in the worst case.

All synchronizing words obtained in a lot of experiments have
lengths near minimal. In particular, no synchronizing word of length
greater than $n^2$ was found.

\subsection{The algorithm for synchronizing word of minimal length}

  The algorithm is a revision of an algorithm for finding
the syntactic semigroup of an automaton
on the base of its transition graph \cite{TW}.
Let us notice that the size of the semigroup is not
polynomial in the graph size and the problem to find synchronizing
word of minimal length is NP-hard \cite{MS}.

We find first some synchronizing word $s$ of length $L$ using 
 mention above algorithms. Foe every left subword $s_i$ of $s$ of length $i$
let us keep the set $\Gamma s_i$ with its size $|\Gamma s_i|$.
The subsets of states of $\Gamma$ are presented by vectors of
units and zeroes, the units correspond to the subset states.

The let us consider the mappings of the graph of the automaton induced
by the letters of the alphabet of the labels are considered.
They correspond to semigroup elements. For every letter $\alpha$
(and a word $t$ $\Gamma \alpha$ (and $\Gamma t$) is a subset of states
can be presented by vector of units and zeroes where the units correspond
to the states of subset.

Let us consider the sequence of these mappings (or vectors). First
vector $v_0$ of the sequence consists of units and presents all states of
$\Gamma$. For every vector $v_i$ of the set of states $S_{v_i}$ from the sequence
we consider the set $S_{v_i} \alpha$ for every letter $\alpha$ of the alphabet
and the corresponding vector $v_i \alpha$.

With every vector $v_i$ we connect the former vector, the letter that has
created it, size of $S_{v_i}$ and the length $l(v_i)$ of the word $u$ such that
$\Gamma u =S_{v_i}$. So for $v_i \alpha$ the former vector is $v_i$,
the letter is $\alpha$ and $l(v_i\alpha)=l(v_i)+1$.

The vector $v_i \alpha$ is excluded from the study if $l(v_i\alpha)>L$ because
the corresponding word could not be a part of a minimal synchronizing word.

We compare the vector $v_i \alpha$ with every vector $w$ of the sequence if 
$l(v_i\alpha)\leq L$. If $S_{w} \subseteq S_{v_i}\alpha$ the vector
 $v_i \alpha$ also is excluded from the study.
 We compare the vector $v_i \alpha$ with every
vector of the set $s_j$ for $j<i$ if $l(v_i\alpha)\leq L$.  
If $S_{s_j} \subseteq S_{v_i}\alpha$ for some $j$ then the vector
 $v_i \alpha$ also is excluded from the study.

 Otherwise, $v_i \alpha$ is added to the sequence. Thus for vector $v_i$
we need $(i-1)d+|s|d$ operations and together with former vectors one has
$i(i-1+|s|)d$ operations. Every vector of the sequence is studied once.

The size of the syntactic semigroup of the automaton is in general not
polynomial in the size of the transition graph. Therefore the time
and space complexity of the algorithm is not polynomial in the size
of the graph in the  worst case.

  With any vector let us connect the previous vector and its letter.
On this way, the path on the graph of the automaton can be constructed.
 Any synchronizing mapping of the set of vertices presents a synchronizing
word. The word can be restored from letters connected with
vectors.

The algorithm founds a list of all words (elements of syntactic
semigroup) of length $k$ where $k$ is growing. The first
synchronizing word of the list is a synchronizing word of the
minimal length.

The algorithm is valid for both complete and non-complete graphs.
The time complexity of the considered procedure is
$O(|\Gamma|d|N|^2)$ with $O(|\Gamma||N|)$ space complexity.
$N$ is here syntactic semigroup of the graph $\Gamma$ over alphabet 
of size $d$.

\section{The algorithm for synchronizing coloring}
The positive solution of the road coloring problem  \cite{Tc}, \cite{Ti}
 is a basis of a polynomial-time implemented algorithm of $O(n^3)$ complexity
in the worst case. The study uses the following theorems.

\begin{thm}
 \label {t2} \emph{\cite{Tc}}
Let every vertex of a strongly connected directed graph $\Gamma$ have the
same number of outgoing edges. Then $\Gamma$ has synchronizing coloring
if and only if the greatest common divisor of lengths of all its cycles
is one.
 \end{thm}

 \begin{thm}
  $\label {ck}$ \emph{\cite{CKK}} \emph{\cite{Tc}}
Let us consider a coloring of an $AGW$ graph $\Gamma$.
Let $\rho$ be the transitive and reflexive closure of the stability
 relation on the obtained automaton.
Then $\rho$ is a congruence relation, $\Gamma/\rho$ is also an
$AGW$ graph and a synchronizing coloring of $\Gamma/\rho$ implies
a synchronizing recoloring of $\Gamma$.
 \end{thm}

The input of the algorithm is a graph with an arbitrary coloring.
The algorithm changes some colors of edges of the graph.
At the end of the work of the implemented version, on the screen appears
the layout of the graph without coloring and then in few seconds of
artificial delay the desired coloring appears.

The work \cite{BP} presented an algorithm of $O(n^2)d$ time complexity.
We still have no information about its implementation.

\subsection{The algorithm for $k$-synchronizing coloring}

A $k$-synchronizing word of a deterministic automaton is a word in
the alphabet of colors at its edges that maps the state set of the automaton
at least on $k$-element subset. A coloring of edges of a
directed strongly connected finite graph of a uniform outdegree
(constant outdegree of any vertex)  is $k$-synchronizing if the coloring
turns the graph into a deterministic finite automaton possessing a
$k$-synchronizing word.

The solution of the problem of $k$-synchronizing coloring based on
 the method from \cite {Tc} appeared first in \cite{BP} and
repeated later independently in \cite{BF}.

Some consequences for coloring of an arbitrary finite digraph
as well as for coloring of such a graph of uniform outdegree are a matter
of the algorithm. The minimal value of $k$ for $k$-synchronizing coloring
is found by the algorithm for any finite digraph. The value of $k$ is equal to
the great common divisor of lengths of cycles of the digraph.
So we obtain a partially synchronizing coloring.

The polynomial-time algorithm for $k$-synchronizing coloring has also $O(n^3)d$
time complexity at worst and quadratic space complexity \cite{Tk}.

\section{The approach to the visualization of digraph}
The visualization of the transition graph of the automaton
is an important help tool of the study of automata.
The visibility of inner structure of a digraph without doubt is a matter of
interest not only for the road coloring, the range of the application may be
significantly wider and includes all directed graphs with labels on edges.

 Crucial role in the visualization plays the correspondence of the layout
to the human intuition, the perception of the structure properties of the
graph and the rapidity of the appearance of the image. The
automatically drawn graphical image must resemble the last one of
a human being and present the structure of the graph. We use and
develop for this goal some known approaches \cite{ST}, \cite{WEK}.

Our main objective is a visual representation of a directed graph with labels
on its edges and, in particular, of the transition
graph of a deterministic finite automaton based on the structure
properties of the graph.
 Among the important visual objects of a digraph one can mention paths,
cycles, strongly connected components, cliques, bunches etc. These
properties reflect the inner structure of the digraph. The pictorial
diagram demonstrates the graph structure highlighting strongly connected
components, paths and cycles. So this kind of visualization can be considered
as a structure visualization.
This algorithm successfully solves a whole series of tasks
of the disposal of the objects.

We choose here a cyclic layout \cite{ST}, \cite{WEK}. According to
this approach the vertices are placed at the periphery of a
circle. Our modification of the approach considered two levels of
circles, the first level consists of strongly connected
components, the second level corresponds to the
 whole graph with $SCC$ at the periphery of the circle. The visual placement
is based on the structure of the graph considered as a union of the set
of strongly connected components.

 Clearly, the curve edges (used, for instance, in the package
GraphViz \cite{EGK}, \cite{Si}) hinder to recognize the cycles and
paths. Therefore, we use only direct and, hopefully, short edges.
We have changed some priorities of the layout and, in particular,
eliminate the goal of reducing the number of intersections of the
edges as it was an important aim in some algorithms \cite{Si}.
The intersections of the edges are even not considered
in our algorithm. This approach gives us an opportunity to
simplify essentially the procedure and to reduce its complexity.
Our main intent is only not to stir by the intersections of the
edges to conceive the structure of the graph. The intersections
are placed in our algorithm far from the vertices due to the
cyclic layout \cite{WEK}, \cite{ST} we use. The area of vertices
differs of the area of the majority of intersections.

The algorithm for the visualization is linear in the size of the
automaton. Thus the linearity of  the algorithm is
comfortably and important.

\subsection{Visualization algorithm}

The layout of the deterministic graph is demonstrated by a high-speed
linear program.

The strongly connected components (SCC) are of special
significance in the algorithm. Thus our first step is the eduction
and selection of the $SCC$. The quick linear algorithm for finding
$SCC$ \cite{AHU} is implemented in the program.

According to the cyclic approach, all $SCC$ are placed on the periphery
of a big circle and are ordered according to the size \cite{TBC}.
The vertices of $SCC$ are arranged in a circle of $SCC$ in the graph layout.
So strongly connected components can be easily recognized by the observer.

The periphery of a circle of $SCC$ is the most desirable area for
placing the edges because the edges in this case are relatively
short. We choose the order of the vertices of the $SCC$ on the
circle according to this purpose. The length of some edges can be
reduced in a such way. It also helps to recognize paths and cycles
on the screen.

From  the other hand, the edges between distinct
$SCC$ are relatively longer than the inner edges of strongly
connected components.

  The problem of the placing of the labels near corresponding edges
is sometimes very complicated and frequently the connection
between the edge and its label is not clear. Our solution is to use colors
on the edges instead of labels and exclude the placing of labels.

The set of loops of arbitrary vertex is placed around the vertex with
some shift that depends on the size of the set. The problem of parallel
edges is solved analogously, the origins of the edges must belong to the vertex.
The complexity of the algorithm shows the following

 \begin{lem} $\label {p4}$
The time and space complexity of the visualization algorithm described above
is linear in the sum of states and edges of the transition graph of the automaton.
 \end{lem}
The transition graph of any deterministic finite automaton is
accepted by the visualization algorithm. The transitions graphs of
non-complete automata also can be reproduced.

\section{Input of data in the package}
The input file is an ordinary txt file for all algorithms used in
the package TESTAS. We open the source file and then check different
properties from menu bar. The graph is shown on the display by
help of a rectangular table. More precisely,
 transition graph of an automaton as well as an arbitrary directed graph
with distinct labels on outgoing edges of every vertex is presented by the matrix
(Cayley graph):

                       \centerline{vertices X labels}

   First two numbers in input file are the size of alphabet of labels
  and the number of vertices.  The integers from 0 to n-1 denote the vertices.
  i-th row is a list of successors of i-th vertex according to the label in the
  column (number of the vertex from the end of  edge with label from the j-th column
  and beginning in i-th state is placed in the (i,j) cell).

  The User defines the data: the number of nodes, size of
 the alphabet of edge labels and the values in the matrix.
 For example, the input 2 6 1 0 2 1 0 3 5 2 3 2 4 5
 presents the Cayley graph with 2 labels and 6 vertices and the next input
  2 5 1 0 2 1 ; 3 5 ; 3 ; presents the  Cayley graph with 2 labels and 5 vertices.
The values are divided by a gap.  The semicolon corresponds to
empty cell of the table.

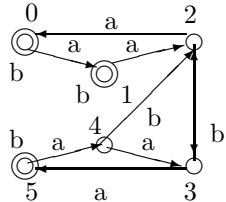
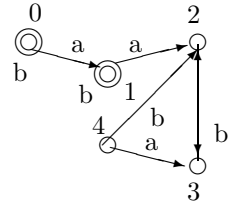
\begin{center}
\begin{tabular}{|c|c|c|}
 \hline
    & letter $ a $ & letter $ b $ \\ \hline
  $ vertex $ 0 & 1 & 0 \\ \hline
  $ vertex $ 1 & 2 &  1 \\ \hline
  $ vertex $ 2 & 0 &  3 \\ \hline
  $ vertex $ 3 & 5 &  2 \\ \hline
  $ vertex $ 4 & 3 & 2 \\ \hline
  $ vertex $ 5 & 4 &  5 \\
  \hline
\end{tabular}  $ and $
\begin{tabular}{|c|c|c|}
 \hline
    & letter $ a $ & letter $ b $ \\ \hline
  $ vertex $ 0 & 1 & 0 \\ \hline
  $ vertex $ 1 & 2 &  1 \\ \hline
  $ vertex $ 2 &  &  3 \\ \hline
  $ vertex $ 3 &  &  \\ \hline
  $ vertex $ 4 & 3 & 2 \\
  \hline
\end{tabular}
\end{center}

\begin{picture}(50,78)
\end{picture}
\begin{picture}(130,78)
\multiput(6,60)(64,0){2}{\circle{6}}
\multiput(6,13)(64,0){2}{\circle{6}}

\put(6,68){0}
 \put(66,67){2}
 \put(42,37){1}
 \put(6,0){5}
 \put(66,0){3}
 \put(30,25){4}

 \multiput(22,56)(22,0){2}{a}
\multiput(16,19)(34,0){2}{a}
 \put(36,21){\circle{6}}
\put(36,48){\circle{6}}
 \put(7,14){\vector(4,1){28}}
\put(7,57){\vector(4,-1){26}}

\put(39,52){\vector(4,1){27}}
 \put(37,20){\vector(4,-1){28}}
\put(67,63){\vector(-1,0){57}}
 \put(36,65){a}
\put(67,12){\vector(-1,0){57}}
 \put(32,0){a}

\put(70,15){\vector(0,1){42}}
 \put(70,59){\vector(0,-1){42}}
\put(34,21){\vector(1,1){36}}
 \put(52,28){b}

  \put(76,22){b}

\put(25,37){b} \put(36,48){\circle{10}} \put(0,20){b}
\put(0,45){b}

\put(6,60){\circle{10}}
 \put(6,13){\circle{10}}
 \end{picture}
\begin{picture}(45,78)
\end{picture}
\begin{picture}(130,78)
\multiput(6,60)(64,0){2}{\circle{6}}
 \put(70,13){\circle{6}}
 \multiput(22,56)(22,0){2}{a}
\put(50,19){a}
 \put(36,21){\circle{6}}
\put(36,48){\circle{6}}

\put(6,68){0}
 \put(66,67){2}
 \put(42,38){1}
 \put(66,0){3}
 \put(30,25){4}

\put(7,57){\vector(4,-1){26}}

\put(39,52){\vector(4,1){27}}
 \put(37,20){\vector(4,-1){28}}

\put(70,15){\vector(0,1){42}}
\put(70,59){\vector(0,-1){42}}
\put(34,21){\vector(1,1){36}}
 \put(52,28){b}

  \put(76,22){b}

\put(25,37){b}
\put(36,48){\circle{10}}

\put(0,45){b}

\put(6,60){\circle{10}}

 \end{picture}
\\
\\
An important verification tool of the package is the possibility
to study the semigroup of an automaton.
The program  finds syntactic semigroup of the automaton,
its size and generators.
The semigroup is presented by a quadratic table of the
form elements X generators (letters). In the cell $(i,j)$ is
a product of element $i$ and generator $j$. The first line of the
table presents the size of the semigroup and the number of generators.

 \end{document}